\documentclass[article]{aa}
\usepackage{graphicx}
\usepackage{natbib}
\usepackage{url}
\usepackage{multirow,bigdelim}
\usepackage{rotating}
\usepackage{placeins}
\bibpunct{(}{)}{;}{a}{}{,} 

\begin{document}
\newcommand{\zabs}{\ensuremath{z_{\rm abs}}}
\newcommand{\avg}[1]{\left< #1 \right>} 
\newcommand{\zem}{\ensuremath{z_{\rm em}}}
\newcommand{\zqso}{\ensuremath{z_{\rm quasar}}}
\newcommand{\zgal}{\ensuremath{z_{\rm gal}}}
\newcommand{\HH}{\mbox{H$_2$}}
\newcommand{\HD}{\mbox{HD}}
\newcommand{\CO}{\mbox{CO}}
\newcommand{\dla}{damped Lyman-$\alpha$}
\newcommand{\Dla}{damped Lyman-$\alpha$}
\newcommand{\lya}{\ensuremath{{\rm Ly}\,\alpha}}
\newcommand{\Lya}{\ensuremath{{\rm Ly}\,\alpha}}
\newcommand{\lyb}{Ly\,$\beta$}
\newcommand{\Ha}{H\,$\alpha$}
\newcommand{\Hb}{H\,$\beta$}
\newcommand{\lyg}{Ly\,$\gamma$}
\newcommand{\lyd}{Ly\,$\delta$}
\newcommand{\ArI}{\ion{Ar}{i}}
\newcommand{\CaII}{\ion{Ca}{ii}}
\newcommand{\CI}{\ion{C}{i}}
\newcommand{\CII}{\ion{C}{ii}}
\newcommand{\CIV}{\ion{C}{iv}}
\newcommand{\ClI}{\ion{Cl}{i}}
\newcommand{\ClII}{\ion{Cl}{ii}}
\newcommand{\CoII}{\ion{Co}{ii}}
\newcommand{\CrII}{\ion{Cr}{ii}}
\newcommand{\CuII}{\ion{Cu}{ii}}
\newcommand{\DI}{\ion{D}{i}}
\newcommand{\FeI}{\ion{Fe}{i}}
\newcommand{\FeII}{\ion{Fe}{ii}}
\newcommand{\GeII}{\ion{Ge}{ii}}
\newcommand{\HI}{\ion{H}{i}}
\newcommand{\MgI}{\ion{Mg}{i}}
\newcommand{\MgII}{\ion{Mg}{ii}}
\newcommand{\MnII}{\ion{Mn}{ii}}
\newcommand{\NaI}{\ion{Na}{i}}
\newcommand{\NI}{\ion{N}{i}}
\newcommand{\NII}{\ion{N}{ii}}
\newcommand{\NV}{\ion{N}{v}}
\newcommand{\NiII}{\ion{Ni}{ii}}
\newcommand{\OI}{\ion{O}{i}}
\newcommand{\OII}{\ion{O}{ii}}
\newcommand{\OIII}{\ion{O}{iii}}
\newcommand{\OVI}{\ion{O}{vi}}
\newcommand{\PII}{\ion{P}{ii}}
\newcommand{\PbII}{\ion{Pb}{ii}}
\newcommand{\SI}{\ion{S}{i}}
\newcommand{\SII}{\ion{S}{ii}}
\newcommand{\SiII}{\ion{Si}{ii}}
\newcommand{\SiIV}{\ion{Si}{iv}}
\newcommand{\TiII}{\ion{Ti}{ii}}
\newcommand{\ZnII}{\ion{Zn}{ii}}
\newcommand{\AlII}{\ion{Al}{ii}}
\newcommand{\AlIII}{\ion{Al}{iii}}
%
\newcommand{\Ho}{\mbox{$H_0$}}
\newcommand{\ang}{\mbox{{\rm \AA}}}
\newcommand{\abs}[1]{\left| #1 \right|} 
\newcommand{\kms}{\ensuremath{{\rm km\,s^{-1}}}}
\newcommand{\cmsq}{\ensuremath{{\rm cm}^{-2}}}
\newcommand{\ergs}{\ensuremath{{\rm erg\,s^{-1}}}}
\newcommand{\ergsa}{\ensuremath{{\rm erg\,s^{-1}\,{\AA}^{-1}}}}
\newcommand{\ergscm}{\ensuremath{{\rm erg\,s^{-1}\,cm^{-2}}}}
\newcommand{\ergscma}{\ensuremath{{\rm erg\,s^{-1}\,cm^{-2}\,{\AA}^{-1}}}}
\newcommand{\msyr}{\ensuremath{{\rm M_{\rm \odot}\,yr^{-1}}}}
\newcommand{\nhi}{n_{\rm HI}}
\newcommand{\fhi}{\ensuremath{f_{\rm HI}(N,\chi)}}
\newcommand{\refs}{{\bf (refs!)}}

\newcommand{\eso}{European Southern Observatory, Alonso de C\'ordova 3107, Vitacura, Casilla 19001, Santiago 19, Chile\label{eso}}
\newcommand{\iap}{Institut d'Astrophysique de Paris, CNRS-UPMC, UMR7095, 98bis boulevard Arago, F-75014 Paris, France\label{iap}}
\newcommand{\iucaa}{Inter-University Centre for Astronomy and Astrophysics, Post Bag 4, Ganeshkhind, 411\,007, Pune, India\label{iucaa}}
\newcommand{\uchile}{Departamento de Astronom\'ia, Universidad de Chile, Casilla 36-D, Santiago, Chile\label{uchile}}
\newcommand{\mpie}{Max-Planck-Institut f\"ur extraterrestrische Physik, Giessenbachstra{\ss}e, D-85748 Garching, Germany\label{mpie}}
\newcommand{\ioffe}{Ioffe Institute, {Polyteknicheskaya 26}, 194021 Saint-Petersburg, Russia\label{ioffe}}

\title{Spotting high-$z$ molecular absorbers using neutral carbon}
\subtitle{Results from a complete spectroscopic survey with the VLT
  \thanks{Based on observations and archival data from the European Southern Observatory (ESO)
    prog. IDs
    060.A-9024,
    072.A-0346,
    278.A-5062,
    080.A-0482,
    080.A-0795,
    081.A-0242,
    081.A-0334,
    082.A-0544,
    082.A-0569,
    083.A-0454,
    084.A-0699,
    086.A-0074 and
    086.A-0643.
    using the Ultraviolet and Visual Echelle Spectrograph (UVES) and X-shooter
at the Very Large Telescope (VLT), on Cerro Paranal, Chile.}
}
\titlerunning{\CI\ as a tracer of molecular gas}

\author{
  P. Noterdaeme\inst{\ref{iap},\thanks{email: noterdaeme@iap.fr}}
\and
C. Ledoux\inst{\ref{eso}}
\and
S. Zou\inst{\ref{iap}}
\and
P. Petitjean\inst{\ref{iap}}
\and
R. Srianand\inst{\ref{iucaa}}
\and
S. Balashev\inst{\ref{ioffe}}
\and
S. L{\'o}pez\inst{\ref{uchile}}
}
\institute{
  \iap 
  \and \eso \and \iucaa \and \ioffe \and \uchile 
}

\authorrunning{P. Noterdaeme et al.}

\abstract{
  While molecular quasar absorption systems provide unique probes of the physical and chemical properties of the gas as well as
  original constraints on fundamental physics and cosmology, their detection
  remains challenging.  Here we present the results from a complete survey for 
  molecular gas in thirty-nine absorption systems selected solely upon the detection of neutral carbon lines in SDSS spectra,
  without any prior knowledge of the atomic or molecular gas content.
  H$_2$ is found in all twelve systems (including seven new detections) where the corresponding lines are covered by the instrument setups and measured
  to have $\log N($H$_2) \gtrsim 18$, indicating a self-shielded regime. We also
  report seven CO detections (7/39) down to $\log N($CO)$\sim 13.5$, including a new one, and put stringent constraints on $N$(CO)
  for the remaining 32 systems. 
  $N($CO) and $N(\CI)$ are 
  found to be strongly correlated with $N($CO)/$N(\CI)\sim$1/10. This suggests that the \CI-selected absorber population is probing gas deeper than
  the \HI-H$_2$ transition in which a substantial fraction of the total hydrogen in the cloud is in the form of H$_2$.
  We conclude that targeting \CI-bearing absorbers is a very efficient way to find 
  high-metallicity molecular absorbers. However, probing the molecular content in lower metallicity regimes as well as
  high column density neutral gas remains to be undertaken to unravel the processes of gas conversion in normal high-$z$ galaxies. }

\date{Received/Accepted}
\maketitle


\section{Introduction}

The detection and analysis of molecular absorption lines along the lines of sight to background light sources has proven
to be an extremely useful tool to investigate the physical and chemical state of the interstellar medium (ISM) thanks to
the sensitive formation, destruction and excitation processes of molecules. Such technique applies from the Solar
neighbourhood towards nearby stars \citep[e.g.][]{Savage77, Boisse13} to the gas in and around high redshift galaxies revealed by
Damped Lyman-$\alpha$ systems (DLAs) \citep[e.g.][]{Levshakov89, Ge97,Petitjean00, Cui05,Srianand05,Noterdaeme08,Jorgenson10,Carswell11,Balashev17}. In addition, the detection of molecular species
at high-redshift provides original and sensitive probes of fundamental
physics and cosmology. Tiny shifts in the relative wavelengths of H$_2$ Lyman and Werner lines can be used to constrain the
possible space-time variation of the proton-to-electron mass ratio down to
a few parts-per-million over a timescale of Gyrs \citep[see][and references therein]{Ubachs16}. The excitation of CO rotational levels provides in turn one of the
best thermometers for measuring the temperature of the cosmic microwave background (CMB) radiation at high redshift
\citep{Srianand08, Noterdaeme11}.
Last but not least, the molecular phase of the ISM makes the link between the gas
accreted onto galaxy and its gravitational collapse that gives birth to stars.
However, the small number of known molecular absorbers contrasts with the huge number of DLAs detected so far
\citep[e.g.][]{Prochaska05,Noterdaeme12c}: only about 25 confirmed high-redshift H$_2$-bearing DLAs have been reported to date
(see \citealt{Balashev17} and references therein), highlighting the small covering factor of the molecular gas and the
need of efficient selection techniques.

 In the local ISM, early works using Copernicus showed that H$_2$ and neutral carbon (\CI) were frequently
  observed in the same absorption systems \citep[e.g.][]{Liszt81}. Despite the high abundance of carbon,
  it is usually found in ionised forms in high-redshift DLAs and the neutral carbon is see only in 
  a small fraction of DLAs that also show H$_2$ absorption 
  \citep[e.g.][]{Ge01,Srianand05}. This is likely due to the first ionisation potential of carbon (11.26~eV) being close 
  to the energy of Lyman-Werner photons that lead to H$_2$ dissociation (through Solomon process, see e.g. 
  \citealt{Stecher67}). \CI\ also conveniently produces absorption lines out of the Lyman-$\alpha$ forest that
  can be identified even at low spectral resolution. 
  We have therefore performed the first blind survey 
  for neutral carbon lines in quasar spectra from the Sloan Digital Sky Survey \citep[][]{Ledoux15}, without any prior
  knowledge of the associated atomic
and molecular content.
The 66 \CI\ candidates constitute our parent sample. We report here on the complete
follow up of this sample with the Ultraviolet and Visual Echelle Spectrograph (UVES) at
  a resolving power $R \sim 50\,000$ and the X-Shooter spectrograph ($R \sim 5000$) at the Very Large Telescope.

\section{Observations and Results}

We obtained spectra for almost all systems 
that are observable from Paranal Observatory, i.e. a sample
of thirty nine confirmed \CI\ absorbers.
Details about the observing procedures, data reduction, and metal line measurements
are presented in Ledoux et al. (in prep). A near-infrared study of the \NaI\ and \CaII\ lines as well as
the dust extinction properties are presented in Zou et al. (submitted). Here, we focus on the detection of H$_2$ and CO.
Wavelengths and oscillator strengths for H$_2$ and CO lines are from the compilations of \citet{Malec10} and \citet{Dapra16}, respectively. 

\subsection{Molecular hydrogen}

We detect H$_2$ absorption lines whenever covered by our spectra (twelve systems). 
Five of these are already reported in the literature \citep{Noterdaeme07,Srianand08,Jorgenson10,Noterdaeme10b,Klimenko16}, from which we have
  taken the H$_2$ column densities,
and seven are new detections.
We estimated
the total H$_2$ column densities for the new detections through Voigt-profile fitting, 
focusing on the low rotational levels that contain most of the H$_2$. 
We note that while the velocity profile of singly ionised metals is wide with a large number of components, we detect H$_2$ only at velocities where \CI\ is also detected. 
  Below we comment on each system, in order of increasing right ascension of the background quasar.

\subsubsection*{J091721+015448, $\zabs=2.107$}
\begin{figure}
  \centering
  \includegraphics[width=\hsize]{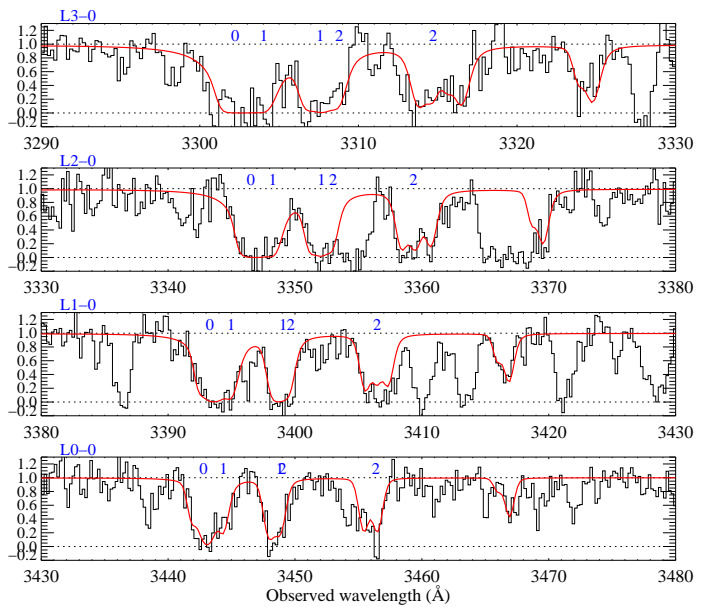}
  \caption{Selected regions of the X-Shooter spectrum of J0917$+$0154 featuring H$_2$ lines. The rotational
    levels J=0 to J=2 are indicated in blue above their corresponding absorption line and the label above
    each panel indicates the band they belong to. Higher rotational levels are fitted but not labelled
  to avoid overcrowding the figure. \label{J0927}}
  \end{figure}

This system was observed with X-shooter at a spectral resolution of $\sim$60~\kms. We obtain an accurate measurement of
the total H$_2$ column density thanks to the damping wings that are seen for the low rotational levels in the four
bands covered by our spectrum (see Fig.~\ref{J0927}) and obtain $\log N($H$_2)=20.11\pm0.06$.

\subsubsection*{J111756+143716, $\zabs=2.001$}

\begin{figure}
  \centering
  \includegraphics[width=\hsize]{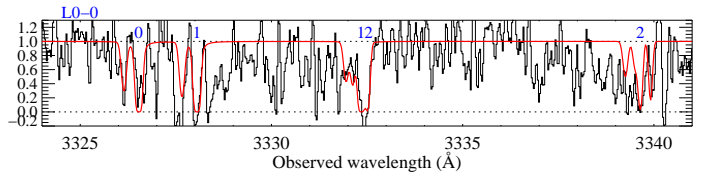}
  \caption{Same as Fig.~\ref{J0927} for the UVES spectrum of J1117$+$1437. \label{J1117}}
\end{figure}

This system is characterised by two narrow H$_2$ components seen in the UVES spectrum (Fig.~\ref{J1117}) in different rotational levels.
  These components also correspond to those seen in the neutral carbon lines. 
While our best-fit value is found to be around $\log N($H$_2)\sim 18$, we note that the data quality is poor and that
only one band is covered, making it impossible to assess the presence of blends. In addition, at such column density,
the absorption is in the logarithmic part of the curve of growth. We are therefore unable to associate an uncertainty to this measurement
that we display with a large arbitrary (albeit quite conservative) 1\,dex error bar in Fig.~\ref{res}.

\subsubsection*{J131129$+$222552, $\zabs=3.092$}

\begin{figure}
  \centering
  \includegraphics[width=\hsize]{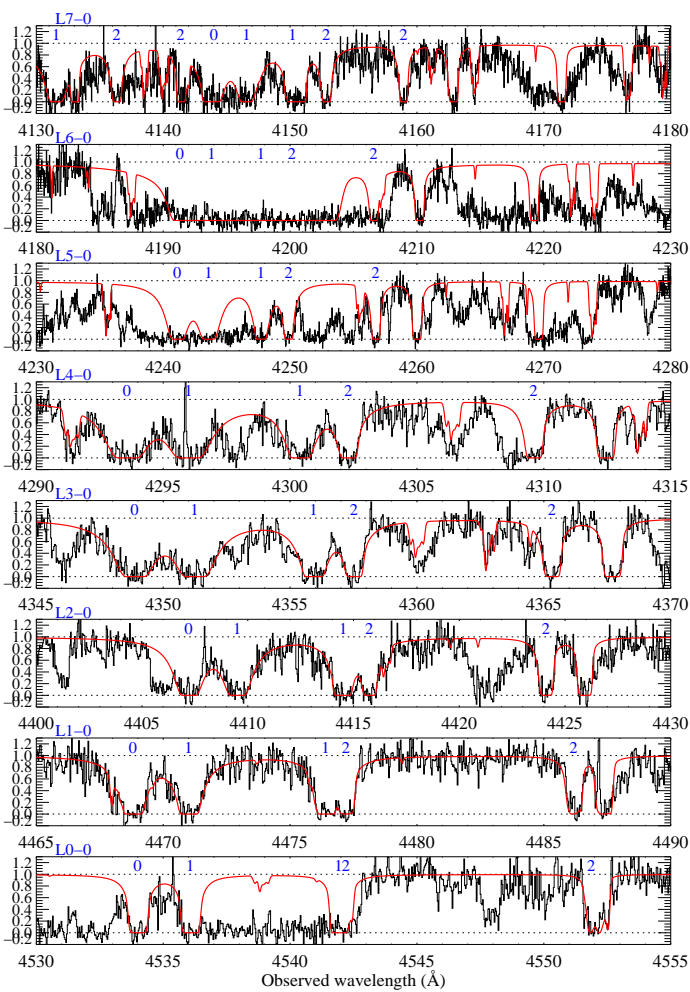}
  \caption{Same as Fig.~\ref{J0927} for the UVES spectrum of J1311$+$2225. Note that different bands start to overlap which
    each other at the shortest wavelength. As for other systems, the label indicated on the top left of each panel
    corresponds to the band to which the identified rotational levels belong. \label{J1311}}
\end{figure}

Thanks to the high absorption redshift, no less than twenty Lyman and Werner H$_2$ bands are covered by our UVES spectrum, shown on Fig.~\ref{J1311}.
Four components can be distinguished in the high rotational levels but lines from the $J=0$ and $J=1$ rotational levels
are strongly damped and therefore modelled using a single component. The damping wings together with the large number
of detected transitions and the high achieved S/N allows a very accurate measurement of the {\sl total} H$_2$ column density which
we found to be $\log N($H$_2)=19.69\pm0.01$.

\subsubsection*{J164610+232922, $\zabs=1.998$}
\begin{figure}
  \centering
  \includegraphics[width=\hsize]{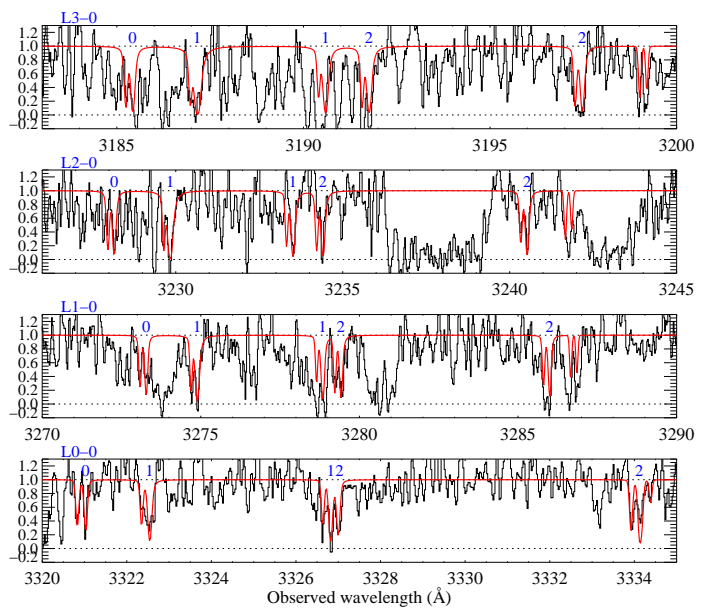}
  \caption{Same as Fig.~\ref{J0927} for the UVES spectrum of J1646$+$2329. \label{J1646}}
\end{figure}

While the S/N of our UVES spectrum in the region of H$_2$ lines (see Fig.~\ref{J1646}) is quite low\footnote{Although SDSS\,J164610+232922 is a relatively
  bright quasar ($g=18.5$), only a single 4000\,s exposure could be obtained at a high airmass (1.7).},
two narrow H$_2$ components are clearly seen in rotational levels J~=~0-3 and our spectrum covers four Lyman bands,
that span more than an order of magnitude in oscillator strengths. We find a total column density $N($H$_2)\approx 10^{18}$~\cmsq\ 
with a $\sim$30\% uncertainty. 

\subsubsection*{J225719$-$100104, $\zabs=1.836$}

\begin{figure}
  \centering
  \includegraphics[width=\hsize]{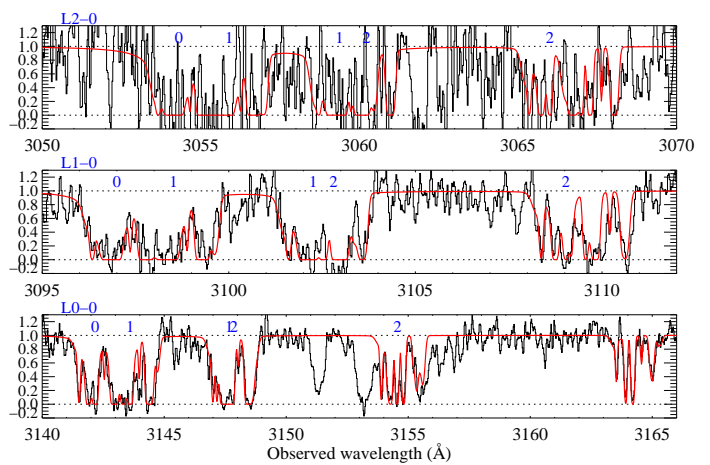}
  \caption{Same as Fig.~\ref{J0927} for the UVES spectrum of J2257$-$1001. \label{J2257}}
\end{figure}

This system is more complex with no less than eight H$_2$ components, strongly blended with each other. Unfortunately, only three Lyman H$_2$ bands are covered by the UVES data (Fig.~\ref{J2257}),
the bluest of which in a region with low S/N ratio. To remove strong degeneracy between parameters, we had 
to fix the excitation temperature $T_{\rm 01}$ 
to 100~K. While this is a strong assumption, we note that varying $T_{\rm 01}$ within a factor of two
has little effect on the total column H$_2$ density (changes $\sim$0.1~dex).
Still, we caution that this error may be underestimated and covering bluer transitions is required to confirm our column density
measurement ($\log N($H$_2) = 19.5 \pm 0.1$). 

\subsubsection*{J233156-090802, $\zabs=2.142$}
\begin{figure}
  \centering
  \includegraphics[width=\hsize]{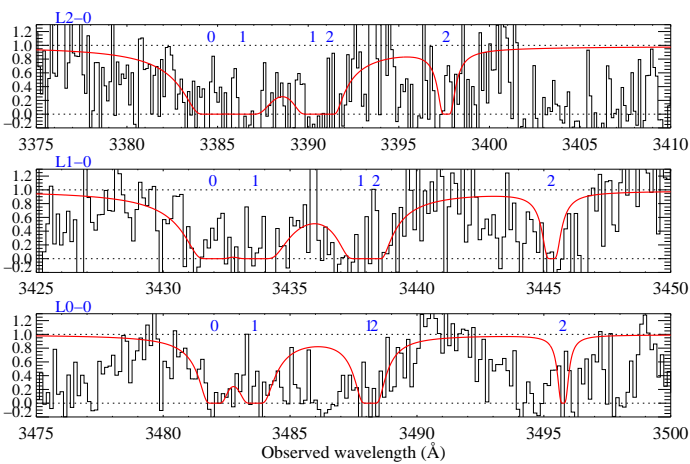}
  \caption{Same as Fig.~\ref{J0927} for the UVES spectrum of J2331$-$0908. The data has been rebinned by 7 pixels for visual purpose only. \label{J2331}}
\end{figure}

In spite of the low S/N achieved this system, shown on Fig.~\ref{J2331}, the data is clearly consistent with strongly damped H$_2$ lines at the same redshift as
that of CO lines (see next section). We fitted the $J=0,1,2$ lines, from which we obtain realistic excitation temperatures,
$T_{01}\sim 140$~K and $T_{02}\sim 180$~K. The total H$_2$ column density is found to be $\log N$(H$_2)=20.57\pm0.05$.

\subsubsection*{J233633-105841, $\zabs=1.829$}
\begin{figure}
  \centering
  \includegraphics[width=\hsize]{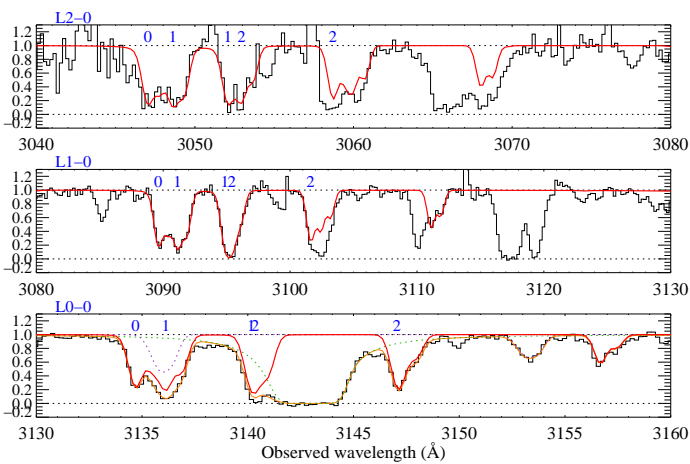}
  \caption{Same as Fig.~\ref{J0927} for the X-Shooter spectrum of J2336-1058. The green and purple dotted lines in the
    bottom panel show the contribution from unrelated Ly$\alpha$ (from a sub-DLA at $\zabs=1.585$) and \OVI\ ($\zabs=2.039$)
    absorption, respectively. The contribution from H$_2$ alone is shown
  in red and the total absorption-line profile is depicted in orange. \label{J2336H2}}
\end{figure}

The H$_2$ profile in this system is well modelled by two components, that are partially blended at the X-Shooter spectral resolution. The bluest
component dominates however the total column density, and the measurement is facilitated by the presence of damping wings and the high S/N
achieved. We note that the L0-0 band is partially blended with unrelated absorption lines, which we modelled when fitting H$_2$ (see Fig.~\ref{J2336H2}). 
We obtain $\log N($H$_2) = 19.0 \pm 0.12$.

\subsection{Carbon monoxide}

CO is detected in seven systems in our sample, six of them already reported by our group and one being
a new detection presented here for the first time. This brings the number of known high-$z$ CO-bearing
quasar absorbers to nine\footnote{The detections towards J1211$+$0833 \citep{Ma15} and J0000$+$0015 \citep{Noterdaeme17} are not formally 
part of the {\sl statistical} sample although selected upon their \CI\ content. \label{foot1}}. 
We measured upper limits on $N$(CO) for all systems assuming the Doppler parameter
to be 1~\kms, similar to what has been measured in all high-$z$ CO absorbers to date. 
We also assume the CMB radiation to be the main excitation source in diffuse gas at high-$z$
\citep[as observed by][]{Srianand08,Noterdaeme11}.

We calculated the local (i.e. for each band individually) and global $\chi^2$ values for a range of
total column densities. 
CO is detected when the $\chi^2$ curves are consistent with each other and present a clear inflexion point, 
defining the best-fit value. For non-detections, $\chi^2(N$(CO)) is generally monotonic with a minimum consistent with that of
$N$(CO)=0 within uncertainty. 
Our 3\,$\sigma$ upper limit corresponds to the column density where the $\chi^2$ is 9 above this minimum. 
With this method, not only we recover all the known CO absorbers but we also identify the new
CO system, at $z=2.143$ towards SDSS\,J2331$-$0908 (Fig.~\ref{CO}), observed by Nestor and collaborators 
(Prog.~ID 080.A-0795). 
This is only the fourth high-$z$ system with direct and simultaneous 
measurements of $N($CO) and $N($H$_2$). 

{Before discussing our findings, it is worth mentioning that, in the local ISM, the excitation temperature of CO is found to be a few degrees above
  the CMB temperature \citep[e.g.][]{Burgh07}, owing to additional excitation processes such as collisions,
  far-infrared dust emission and possibly cosmic rays. Relaxing our assumptions we find that the derived CO column density limits
  (as well as the CO column density for the new detection at $\zabs=2.143$ towards SDSS\,J2331$-$0908) are not changed significantly as the total band equivalent width is almost conserved. For example, allowing an
  excitation temperature 5\,K above the CMB temperature only increases the derived values by less than 0.04~dex.

\begin{figure}
  \centering
  \setlength{\tabcolsep}{1pt}
\renewcommand{\arraystretch}{0.2}
  \begin{tabular}{cc}
    \includegraphics[bb=220 150 414 565,clip=,angle=-90,width=0.48\hsize]{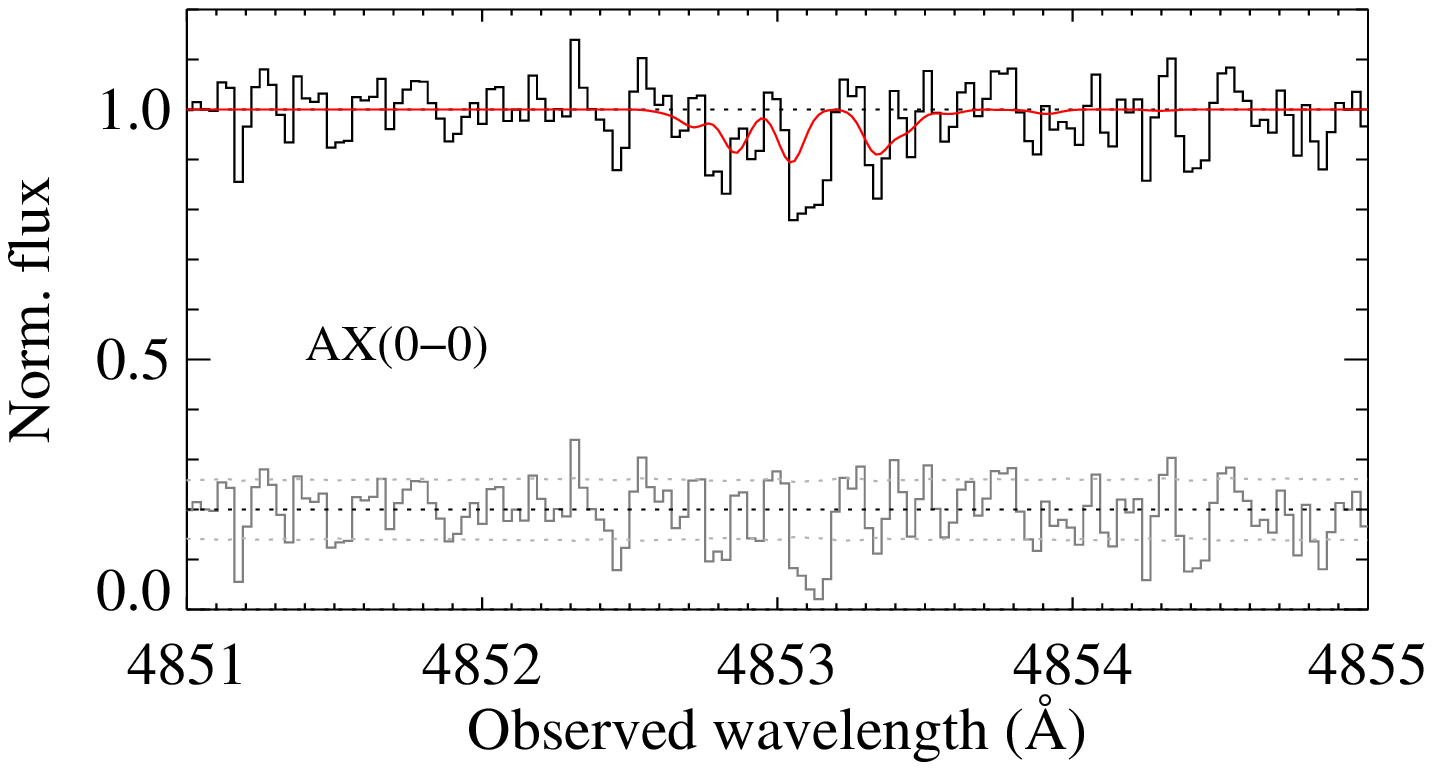} &
    \includegraphics[bb=220 150 414 565,clip=,angle=-90,width=0.48\hsize]{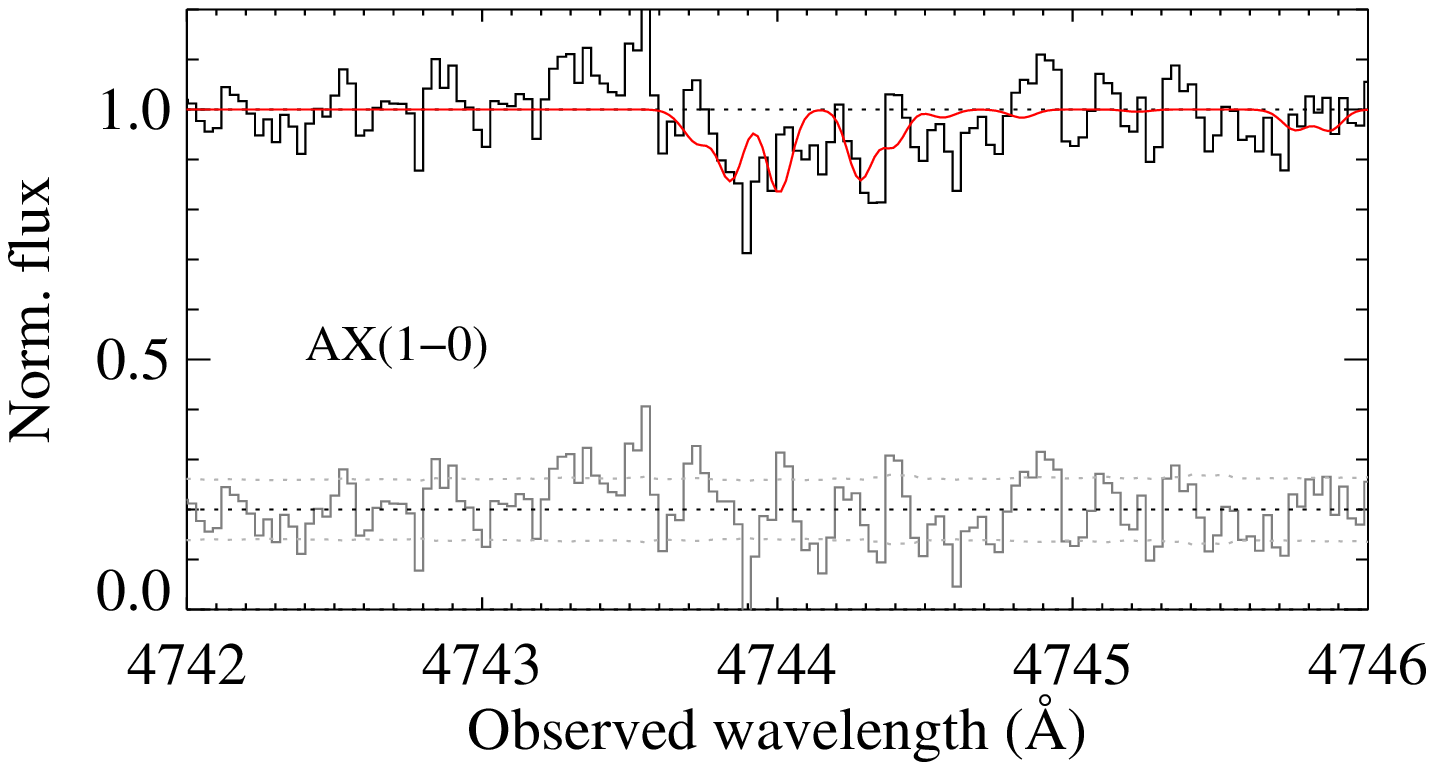} \\
    \includegraphics[bb=220 150 437 565,clip=,angle=-90,width=0.48\hsize]{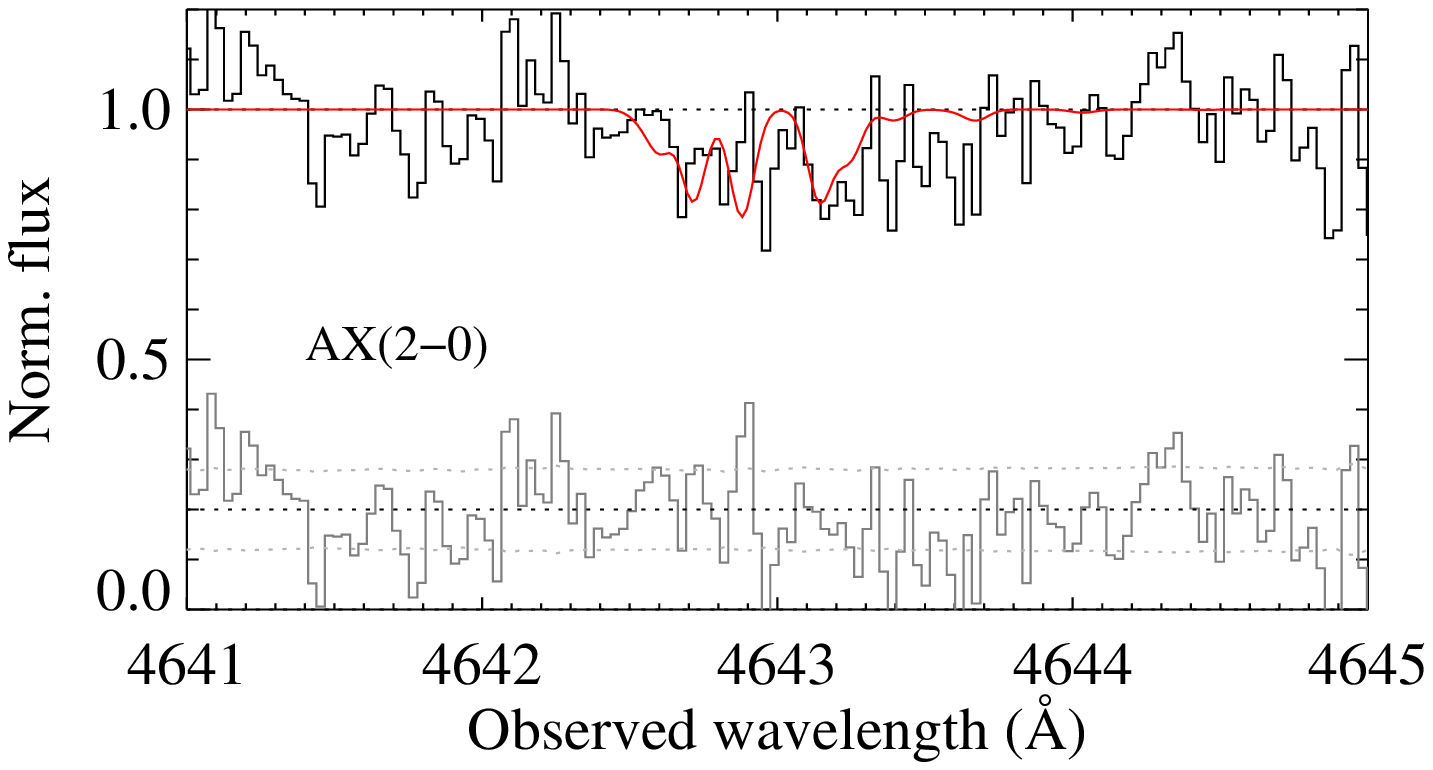} &
    \includegraphics[bb=220 150 437 565,clip=,angle=-90,width=0.48\hsize]{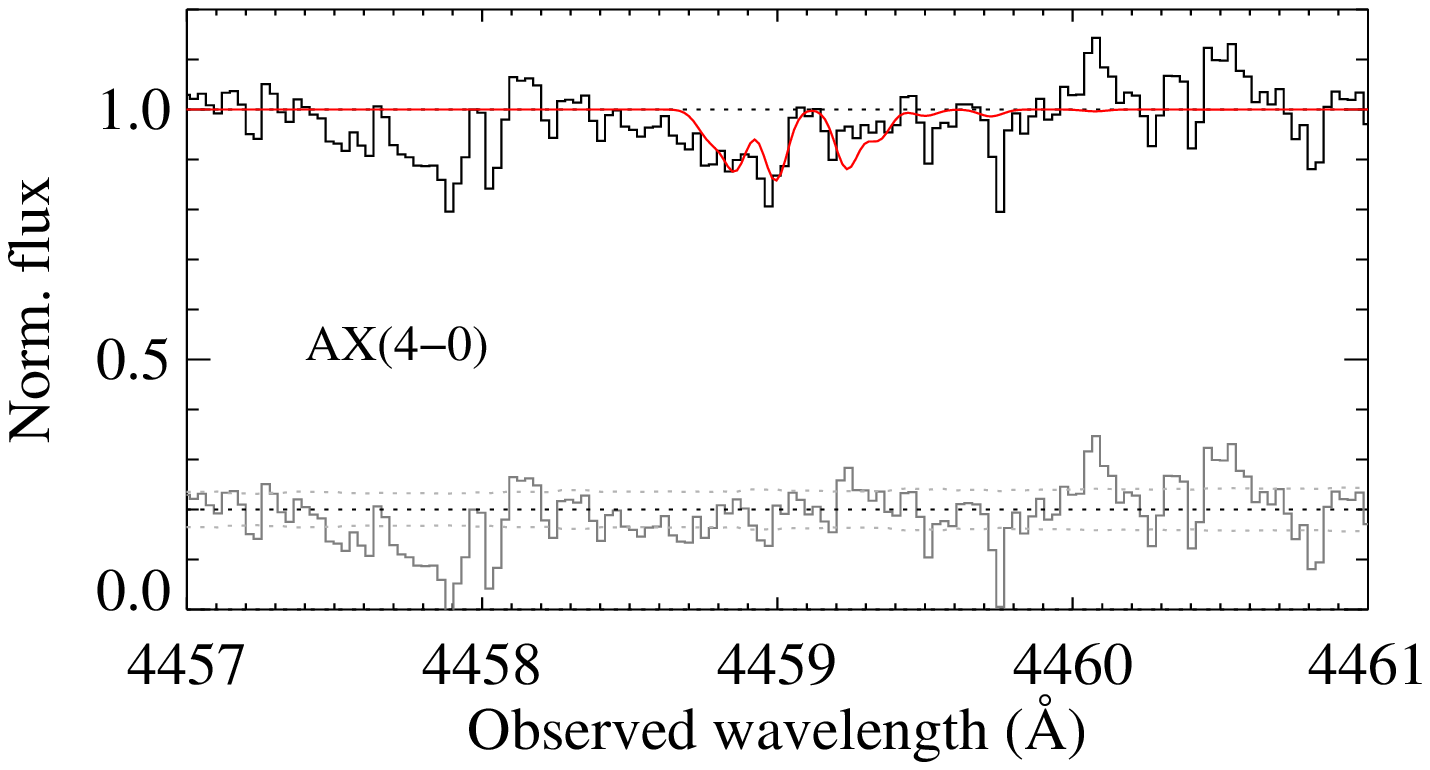} \\
     \includegraphics[bb=170 225 402 640,clip=,angle=90,width=0.48\hsize]{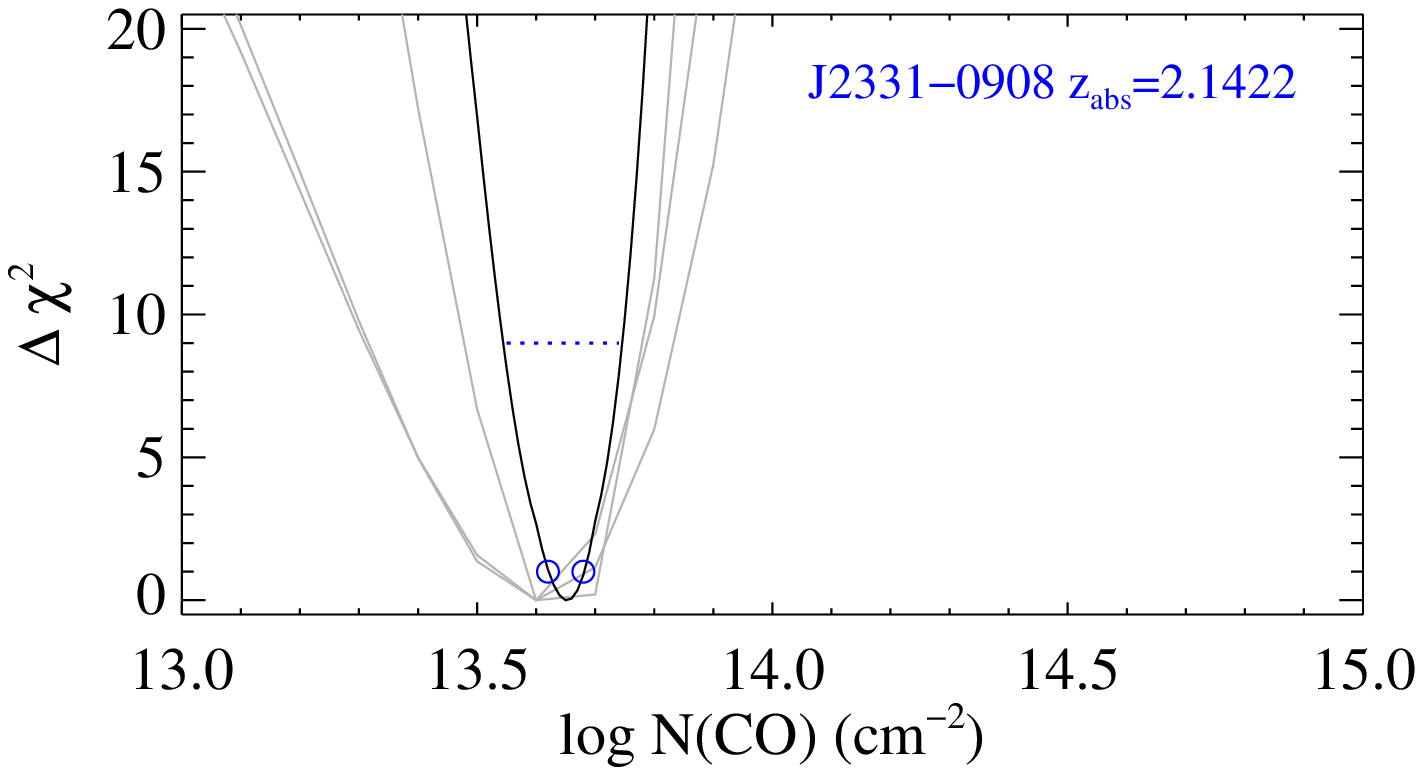} &
     \includegraphics[bb=170 230 402 645,clip=,angle=90,width=0.48\hsize]{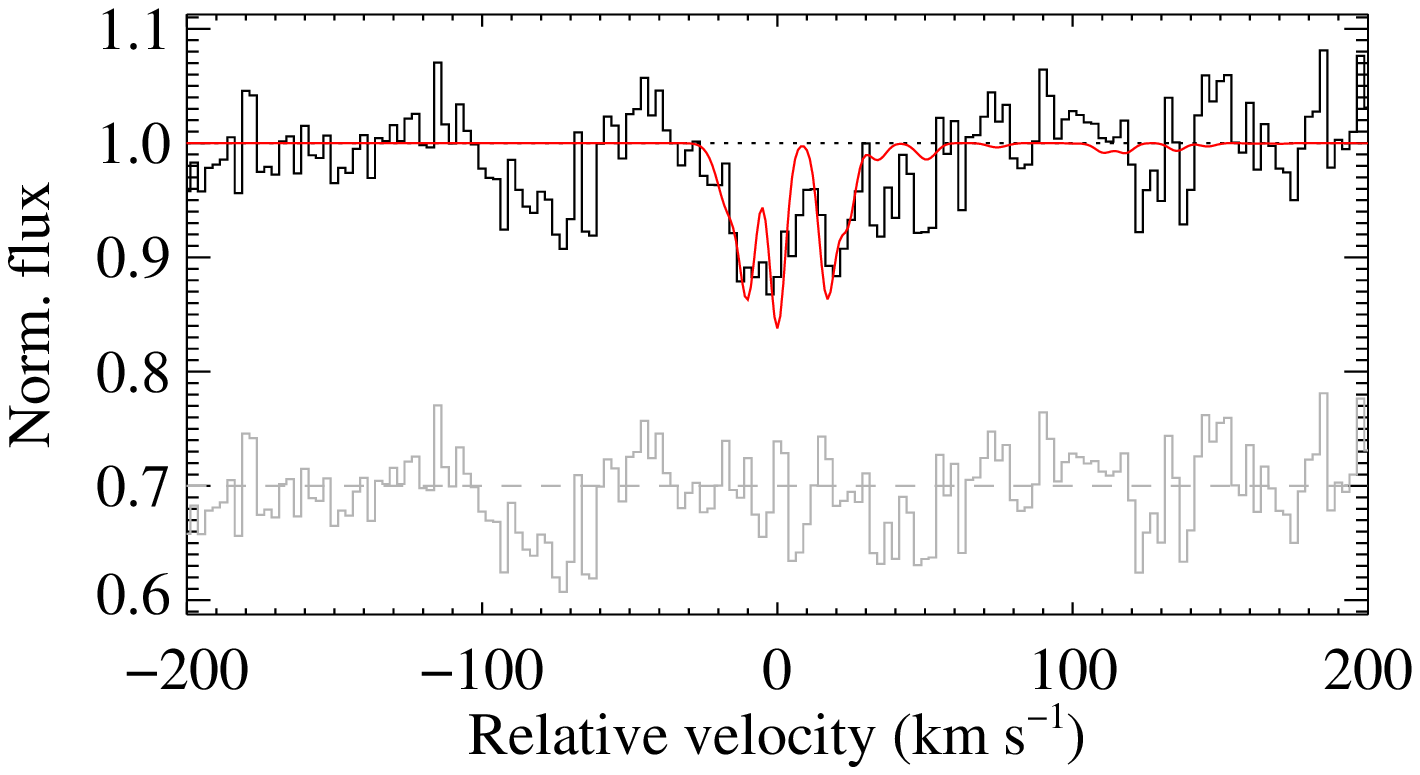} \\
  \end{tabular}
  \caption{CO AX bands at $\zabs=2.1422$ towards J\,2331$-$0908 (top four panels, AX(3-0) is not covered by the instrument setup).
    The bottom-right panel shows a co-addition of the CO bands, using $(f/\sigma)^2$-weighting, where $f$ is the oscillator strength
      and $\sigma$ the uncertainty on the normalised flux, 
      for easy visualisation of the detection. 
    The bottom-left panel shows the $\Delta \chi^2 = \chi^2-\chi^2_{\min}$ curves (grey for individual band, black for total).
    \label{CO}}
\end{figure}

\begin{table}
  \centering
    \setlength{\tabcolsep}{2pt}
  \caption{CO and H$_2$ content of strong C\,{\sc i} absorbers \label{tab}}
\begin{tabular}{c c c c c c c}
  \hline \hline
  Quasar         &$z_{\rm abs}$&  \multicolumn{3}{c}{$\log N$ (\cmsq)} & Ref \\
  &           &  \CI &    CO             &  H$_2$ \\            
  \hline                                               
J0216$-$0021 &  1.737   & 14.25$\pm$0.02 &  $<$12.97         &  --               &  1 \\
J0300$-$0721 &  1.536   & $>$14.77       &  $\le$13.79       &  --               &  1\\
J0811$+$0838 &  1.906   & 13.68$\pm$0.13 &  $<$13.24         &  --               &  1\\
J0815$+$2640 &  1.681   & $>$14.73       &  $<$13.68         &  --               &  1\\
J0820$+$1559 &  1.547   & $>$14.71       &  $<$14.01         &  --                 &  1\\
J0852$+$1935 &  1.788   & 15.01$\pm$0.12 &  $<$13.33         &  --               &  1\\
J0854$+$0317 &  1.567   & 14.23$\pm$0.01 &  $<$13.11         &  --               &  1\\
J0857$+$1855 &  1.730   & 14.57$\pm$0.13 &  13.54$\pm$0.05   &  --               &  2 \\
J0917$+$0154 &  2.107   & 14.32$\pm$0.06 &  $<$14.07         &$20.11^{+0.06}_{-0.06}$     &  1\\
J0927$+$1543 &  1.731   & $>$14.61       & $<$13.32         &  --               &  1\\     
J1047$+$2057 &  1.775   & $>$14.90       &  14.40$\pm$0.07    &  --               &  2 \\
J1117$+$1437 &  2.001   & 14.40$\pm$0.03 &  $<$13.13         & $\sim 18$         &  1\\   
J1122$+$1437 &  1.554   & 13.83$\pm$0.03 &  $<$12.99         &  --               &  1\\   
J1129$-$0237 &  1.623   & $>$14.96       &  $<$13.31         &  --               &  1\\   
J1133$-$0057 &  1.706   & 15.12$\pm$0.06 &  $\le$14.07       &  --               &  1\\   
J1237$+$0647 &  2.691   & 15.01$\pm$0.02 &  14.17$\pm$0.09   &$19.21^{+0.13}_{-0.12}$ & 3 \\
J1248$+$2848 &  1.513   & 14.25$\pm$0.10 &  $<$13.25         &  --                &  1\\
J1302$+$2111 &  1.656   & 14.30$\pm$0.02 &  $<$13.48         &  --                &  1\\
J1306$+$2815 &  2.012   & 14.47$\pm$0.04 &  $<$13.26         &  --                &  1\\
J1311$+$2225 &  3.092   & 14.30$\pm$0.02 &  $<$13.43         &$19.69^{+0.01}_{-0.01}$ & 1 \\
J1314$+$0543 &  1.583   & 14.40$\pm$0.02 &  $\le$13.77       &  --                &  1\\   
J1341$+$1852 &  1.544   & 13.51$\pm$0.03 &  $<$13.00         &  --                &  1\\
J1346$+$0644 &  1.512   & 14.51$\pm$0.02 &  $<$13.60         &  --                &  1\\
J1439$+$1117 &  2.418   & 14.81$\pm$0.02 &  13.89$\pm$0.02   &$19.38^{+0.10}_{-0.10}$ &  4 \\
J1459$+$0129 &  1.623   & 14.32$\pm$0.09 &  $<$13.57         &--                   &  1\\   
J1522$+$0830 &  1.627   & $>$14.47       &  $<$13.50         &--                   &  1\\   
J1603$+$1701 &  1.890   & 13.80$\pm$0.10 &  $<$12.88         &--                   &  1\\   
J1604$+$2203 &  1.641   & $>$15.14       &  14.59$\pm$0.11   &--                   &  5\\    
J1615$+$2648 &  2.118   & 14.49$\pm$0.06 &  $<$13.16         &--                   &  1\\    
J1623$+$1355 &  1.751   & 14.41$\pm$0.07 &  $<$13.30         &--                   &  1\\
J1646$+$2329 &  1.998   & 14.32$\pm$0.06 &  $<$13.40         & $18.02^{+0.11}_{-0.11}$ &  1\\
J1705$+$3543 &  2.038   & $>$15.01       &  14.14$\pm$0.03   &--                   &  2\\
J2123$-$0050 &  2.060   & 14.11$\pm$0.02 &  $<$13.07         &$17.94^{+0.01}_{-0.01}$  &  1,6\\
J2229$+$1414 &  1.586   & 13.96$\pm$0.05 &  $<$13.55        &--                    &  1\\
J2257$-$1001 &  1.836   & 14.65$\pm$0.01 &  $<$13.09         &$19.5 \pm 0.1$       &  1\\
J2331$-$0908 &  2.143   & $>$14.70       &  13.65$\pm$0.03   &$20.57^{+0.05}_{-0.05}$  &  1\\
J2336$-$1058 &  1.829   & 14.07$\pm$0.02 &  $<$12.93         &$19.00^{+0.12}_{-0.12}$  &  1\\
J2340$-$0053 &  2.054   & 13.99$\pm$0.02 &  $<$12.58         &$18.47^{+0.04}_{-0.04}$  &  1,7\\
J2350$-$0052 &  2.426   & 14.36$\pm$0.01 &  $<$12.94         &$18.52^{+0.29}_{-0.49}$  & 1,8 \\
\hline
\end{tabular}

\tablebib{The references listed in last column are for molecular measurements.
  When two references are listed, they correspond to CO then H$_2$, in this order.
(1)~This work; (2) \citet{Noterdaeme11}; (3) \citet{Noterdaeme10b}; (4) \citet{Srianand08};
  (5) \citet{Noterdaeme09}; (6) \citet{Klimenko16}; (7) \citet{Jorgenson10}; (8) \citet{Noterdaeme07}.
Unless already available from the literature, \CI\ column densities
 were obtained through the apparent optical depth method.
}
\end{table}

\section{Discussion}

\begin{figure}
  \centering
  \renewcommand{\arraystretch}{0.5}
  \begin{tabular}{c}
    \includegraphics[width=0.9\hsize,bb=70 222 485 464,clip=]{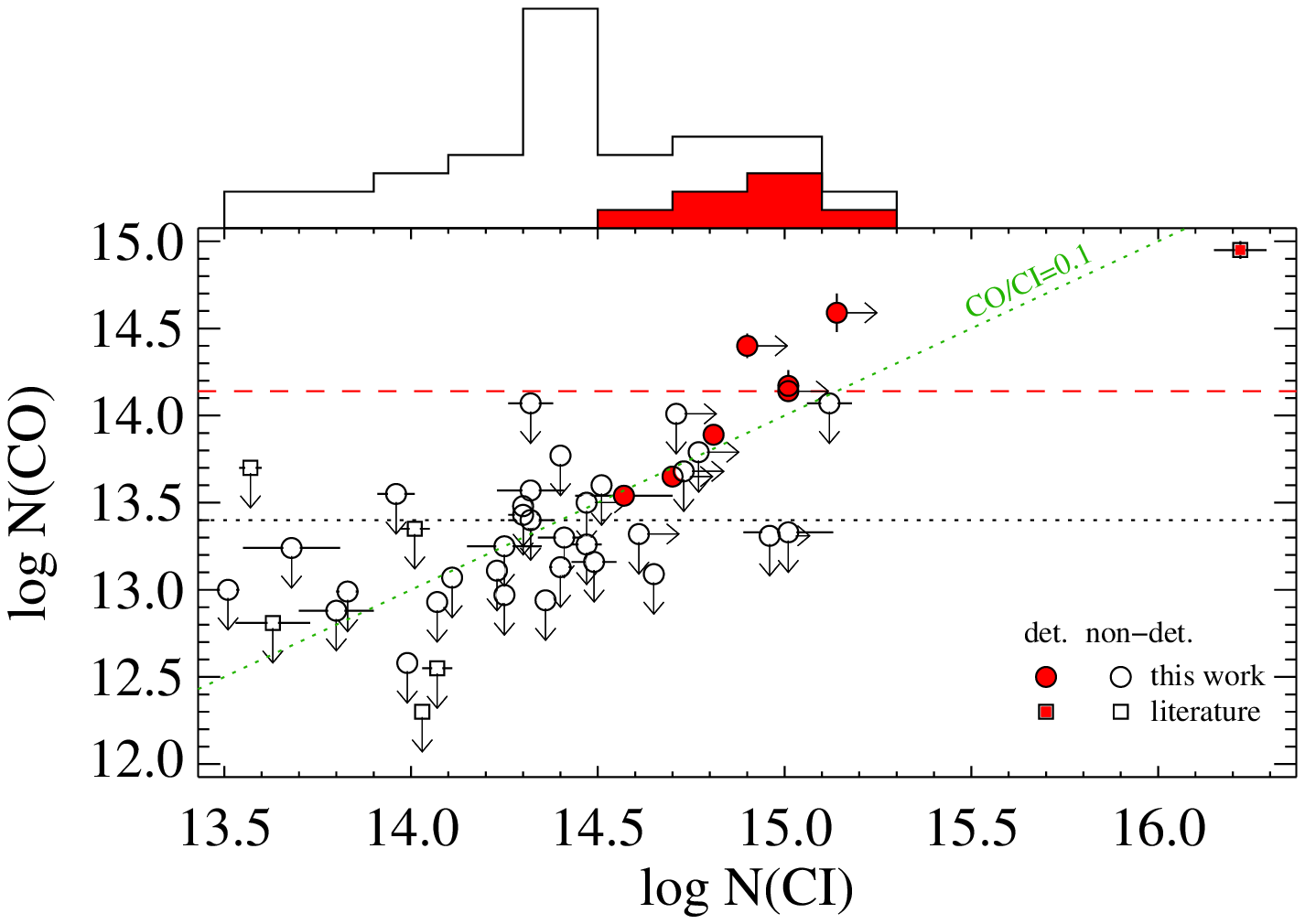}\\
    \includegraphics[width=0.9\hsize,bb=70 222 485 397,clip=]{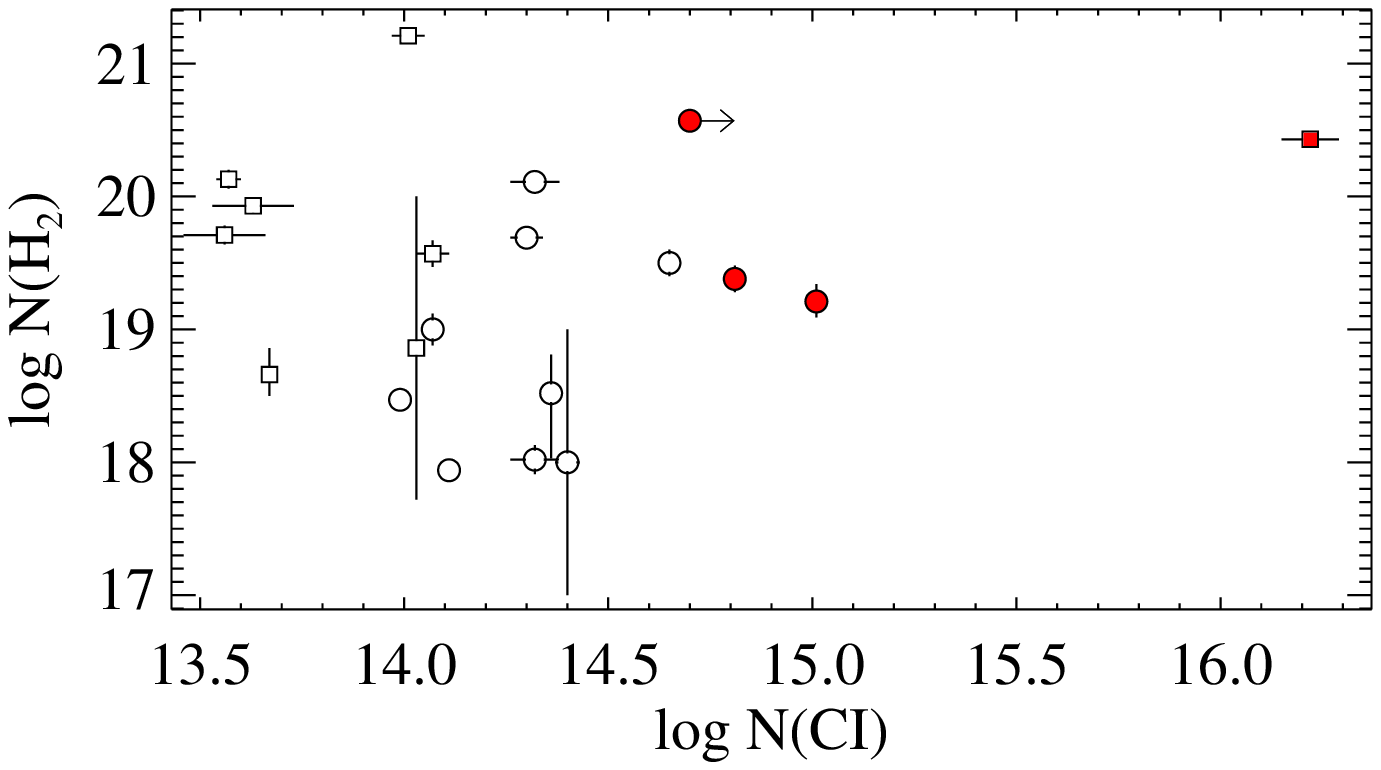}\\
    \includegraphics[width=0.9\hsize,bb=70 176 485 397,clip=]{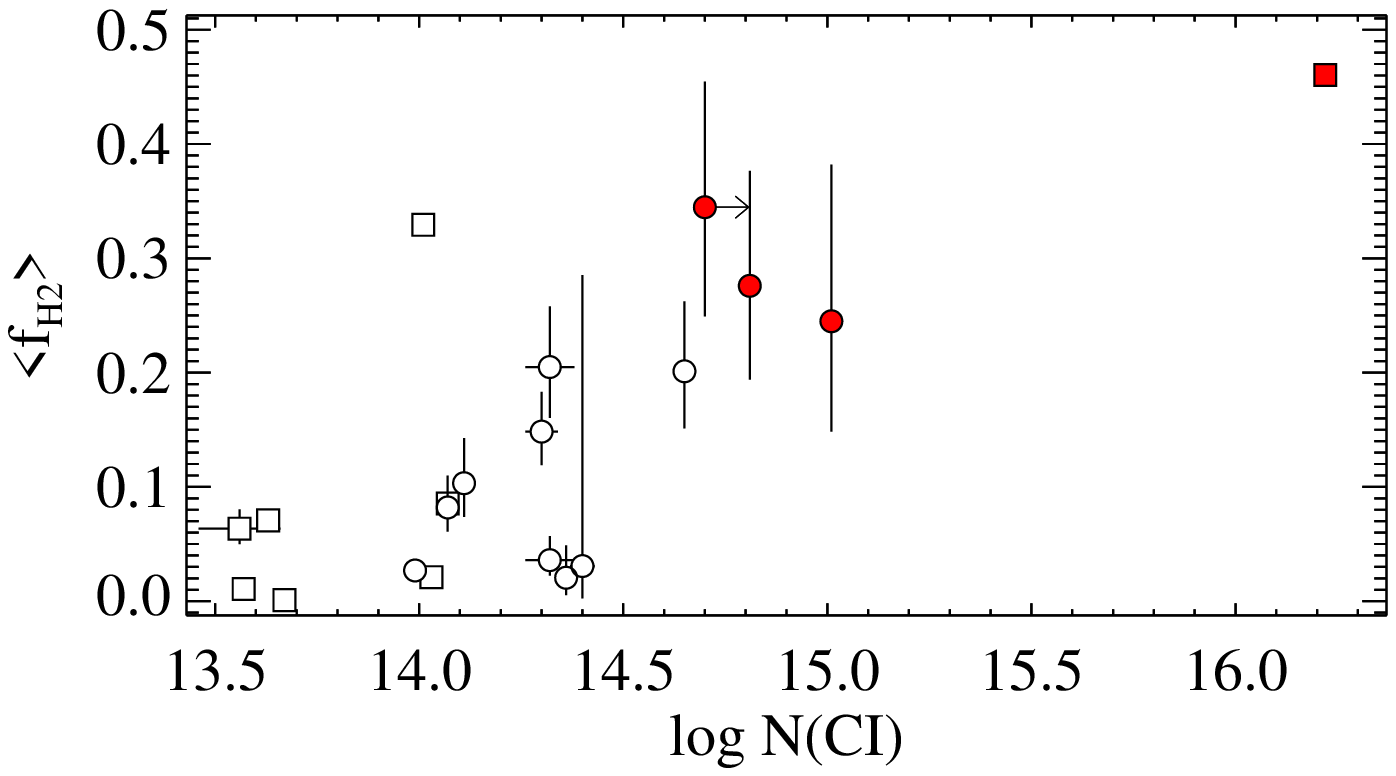}
  \end{tabular}
  \caption{Column densities of CO (top), H$_2$ (middle) and overall molecular fractions (bottom) vs $N(\CI)$.
    CO detections are represented by red colours. The $N(\CI)$-distributions and median $\log N($CO) values (horizontal lines) 
    are shown for the statistical sample only (circles).  
    Squares correspond to high-$z$ H$_2$ DLA systems from the literature \citep{Balashev10,
      Balashev11,Balashev17,Carswell11,Guimaraes12,Noterdaeme15,Noterdaeme17,Petitjean02}.\label{res}}
\end{figure}

Table~\ref{tab} summarises the H$_2$ and CO detections and column density measurements.
Figure~\ref{res} presents the H$_2$ and CO column densities as well as the {\sl overall} molecular fraction $\avg{f_{\rm H2}}=2N($H$_2)/(2N$(H$_2)+N(\HI))$
as a function of $N(\CI)$ for our complete sample.
Known systems from the literature are
also added for comparison but not considered for statistical analysis.

We find that H$_2$ is detected with $N($H$_2) \ga 10^{18}$~\cmsq\ in all systems with $\log N(\CI)>13.5$. In this
regime, H$_2$ is likely to be self-shielded and the molecular fraction substantial in the H$_2$-bearing gas.
We also observe a possible trend for increasing $N$(H$_2$) with increasing $N(\CI)$ (Spearman rank correlation coefficient $r=0.4$, 1.2\,$\sigma$ significance) in our statistical sample, albeit with a large dispersion. We note that systems that were {\sl not} \CI-selected (from literature)
seem not to follow this trend. Four of them indeed have $N$(H$_2) > 5\times 10^{19}$\,\cmsq\ in spite of relatively low $\CI$ column density ($\log N(\CI)\la 14$).
This difference is likely due to the different chemical enrichments: \CI-selected systems probe mostly high-metallicity gas
(Zou et al. sub., Ledoux et al. in prep.) while the four mentioned literature systems all have low metallicities.

Since the column density at which \HI\ is converted into H$_2$ strongly depends on the chemical
properties of the gas, in particular the abundance of dust grains \citep[e.g.][]{Bialy16}, we can expect less \HI\ in the molecular
cloud envelope for high-metallicity systems compared to low-metallicity ones.
In addition, contrary to DLAs, \CI\-systems were selected without
any a priori knowledge of the \HI\ content \citep{Ledoux15} and should have less contribution from unrelated atomic gas that does not belong to the envelope of the H$_2$ cloud. This is seen in the 
bottom panel of Fig.~\ref{res}, where
the correlation between $\avg{f}$ and $N(\CI)$ is seen with $r=0.6$ at 2.1\,$\sigma$: the average
overall H$_2$ molecular fraction is about 15\% in our sample (and about 30\% when CO is detected) but $<3\%$ at $\log N(\CI)<14$.

The correlation between $N($CO$)$ and $N$(\CI) for CO detections is strong with $r=0.88$ (2.6\,$\sigma$).
From the column density distributions, we can see that the probability to detect CO becomes much larger
above $N(\CI)\sim 5\times 10^{14}$~\cmsq (6/12) than below this value (1/27). In addition, there is no CO detection among the 18 systems with
$\log N(\CI)<14.4$. Since the CO detection limits are significantly below ($\sim$ 1 dex) the typical $N($CO) in the case of detection, this result
is robust\footnote{We note that $N$(CO)-limits also tend to be more stringent for systems with low $N(\CI)$.
  This is likely due to easier detection of
  small \CI\ equivalent widths in SDSS towards bright quasars (or that such systems extinguish less the same), for which the follow-up data is also of better quality.}.
Considering also lower and upper limits on both $N($CO) and $N(\CI)$, we still find the $N($CO$)$-$N$(\CI) correlation to have
$\sim$92\% probability.

This strong correlation is likely due to the strong dependence of CO abundance on the metallicity (through the abundance of carbon,
the abundance of dust grains as molecule-formation catalyst and an effective shielding of UV photons). In \citet{Ledoux15}, we showed that
strong \CI\ systems produce more reddening than other classes of quasar absorbers. We further note that the dust-reddening is systematically higher
in CO-bearing systems than other H$_2$ systems without CO.
The relative behaviours of CO, H$_2$ and \CI\ agree qualitatively with models of ISM clouds immersed into a UV radiation field:
these clouds are expected to exhibit an onion-like structure where hydrogen converts from atomic to molecular form when going towards the centre of
the cloud. Carbon is predominantly ionised in the external layers, then becomes neutral, while CO is dominant only in the inner dense molecular parts
of the cloud (see. e.g. \citealt{Bolatto13}). Unfortunately, it remains difficult to disentangle the atomic gas that belongs to a molecular cloud envelope
and contributes to its shielding from unrelated \HI, simply intercepted along the same line of sight. This means that the measured $\avg{f}$ 
is a lower limit to the actual H$_2$ molecular fraction in the \CI-bearing cloud.
Since CO and \CI\ are only found in shielded gas, their observed abundance ratio should be less affected by the presence of unrelated gas. Indeed, we find CO/\CI$\sim 0.1$
for all detections (green dotted line in Fig.~\ref{res}), a value which is also consistent with the non-detections at lower $N(\CI)$.  This indicates a regime
deeper than the layer where the \HI-H$_2$ transition occurs.

The CO/H$_2$ ratio is found to be low ($\sim [3-9]\times 10^{-6}$) for three out of four cases where both these molecules are detected and can be more than an order
of magnitude lower in other strong H$_2$ systems, including the new CO detection. Even in these cases, the high $N($H$_2$) likely indicates well-molecularized regions.
Several factors such as the grain size distribution or the intensity of the cosmic ray field likely play important roles in determining whether CO will be present
or not in H$_2$-dominated regions \citep[e.g.][]{Shaw16,Noterdaeme17, Bisbas17}. 
Multiple clouds can also easily explain large H$_2$ column densities
without significant CO, in a similar way than multiple \HI-H$_2$ transition layers explain higher $N(\HI)$ than predicted by single cloud models \citep{Bialy17}.

We conclude that \CI\ is a very good proxy to spot high-redshift molecular absorbers that can be used for
a variety of studies including fundamental physics and cosmology. It is however crucial not only to constrain the physical parameters in
individual systems (and hopefully for individual velocity components separately) but also to explore different metallicity regimes \citep{Balashev17} using different
selections \citep[e.g.][]{Balashev14} to understand better the molecular structure of ISM clouds at high redshifts.

\begin{acknowledgements}
  We thank T. Kr\"uhler for help with the X-shooter data reduction.
  PN thanks the European Southern Observatory for hospitality and support during part of this work was done.
  PN, PPJ and RS acknowledge support from
  the Indo-French Centre for the Promotion of Advanced Research (Project 5504-B). We acknowledge support from
  the PNCG funded by CNRS/INSU-IN2P3-INP, CEA and CNES, France.
  This research is part of the projet {\sl HIH2} funded by the 
  {\sl Agence Nationale de la Recherche}, under grant ANR-17-CE31-0011-01 (JCJC). 
  SB thanks the Institut d'Astrophysique de Paris for hospitality and the Institut Lagrange de Paris for financial support. 
  SL has been supported by FONDECYT grant 1140838 and by PFB-06 CATA.
\end{acknowledgements}

\bibliographystyle{aa} 
\bibliography{CIbib} 

\end{document}